
\input phyzzx.tex
\date{February 24, 1995}
\voffset=3pc
\hoffset=0.45in

\def\PL  #1 #2 #3 {{\sl Phys.~Lett.}~{\bf#1} (#3) #2 }
\def\NP  #1 #2 #3 {{\sl Nucl.~Phys.}~{\bf#1} (#3) #2 }
\def\PR  #1 #2 #3 {{\sl Phys.~Rev.}~{\bf#1} (#3) #2 }
\def\PRE  #1 #2 #3 {{\sl Phys.~Reports}~{\bf#1} (#3) #2 }
\def\PRD #1 #2 #3 {{\sl Phys.~Rev.~D} {\bf#1} (#3) #2 }
\def\PRB #1 #2 #3 {{\sl Phys.~Rev.~B} {\bf#1} (#3) #2 }
\def\PP  #1 #2 #3 {{\sl Phys.~Rep.}~{\bf#1} (#3) #2 }
\def\MPL #1 #2 #3 {{\sl Mod.~Phys.~Lett.}~{\bf#1} (#3) #2 }
\def\CMP #1 #2 #3 {{\sl Comm.~Math.~Phys.}~{\bf#1} (#3) #2 }
\def\PRL #1 #2 #3 {{\sl Phys.~Rev.~Lett.}~{\bf#1} (#3) #2 }
\def\TMP  #1 #2 #3 {{\sl Theor.~Math.~Phys.}~{\bf#1} (#3) #2 }
\def\JMP  #1 #2 #3 {{\sl Jour.~Math.~Phys.}~{\bf#1} (#3) #2 }
\def\JSP  #1 #2 #3 {{\sl Jour.~Stat.~Phys.}~{\bf#1} (#3) #2 }
\def\IJ  #1 #2 #3 {{\sl Int.~Jou.~Mod.~Phys.}~{\bf#1} (#3) #2 }
\REF\dyson{F. Dyson, \CMP 12 91 1969 ; G. Baker, \PRB 5 2622 1972 . }
\REF\wilson{K.~Wilson, \PRB 4 3185 1971 ; K. Wilson
and J. Kogut, {\sl Phys. Rep.}
{\bf 12} (1974) 75  .}
\REF\sinai{P. ~Bleher and Y. ~Sinai, \CMP 45 247 1975 ; P.~Collet and
J. P. ~Eckmann,\CMP 55 67 1977  and {\it
Lecture Notes in Physics} {\bf 74} (1978) .}
\REF\koch{H. Koch and P. Wittwer, \CMP 106 495 1986  , {\bf 138}  (1991) 537 ,
{\bf 164} (1994) 627  .}
\REF\gaw{K. Gawedzki
and A. Kupiainen, Les Houches 1985, K. Osterwalder and R. Stora, Editors }
\REF\marseille{Y. Meurice, in {\it
Proceedings of the International Europhysics
Conference on High Energy Physics}, p. 89, Eds. J. Carr and M. Perottet,
Editions Fronti\`eres, 1994.}
\REF\high{Y. Meurice, \JMP 36 000 1995 (in press),
U. of Iowa preprint 94-14 hep-lat 9408021.}
\REF\num{Y. Meurice, G. Ordaz and V.G.J. Rodgers, \JSP 77 607 1994 .}
\REF\epsi{P. Collet, J.-P. Eckmann, and B.
Hirsbrunner, {\it Phys. Lett.} {\bf 71B}, 385 (1977). }
\REF\gaunt{D. S. Gaunt and
A. J. Guttmann, in {\it Phase Transitions and Critical Phenomena},
(C. Domb and M. S. Green, eds.) Academic Press 1974. }
\REF\creutz{G. Bhanot,
M. Creutz, U. Glasser and K. Schilling, {\it Phys. Rev. B } {\bf 49}
, 12909 (1994) . }
\REF\fisher{M. Fisher and D. Gaunt \PR 133A 224 1964 .}
\REF\veneziano{G. Veneziano, \PL B124 357 1983 ; for
more references see: D. Amati, K. Konishi,
Y. Meurice, G. C. Rossi and G. Veneziano, \PRE 162 169 1988 .}
\REF\rw{Y.Meurice, \PL 265B 377 1991 ; J.L.Lucio and Y. Meurice,
{\sl Mod. Phys. Lett.} {\bf 6} (1991) 1199 . }
\REF\conje{Y. Meurice, \MPL 7 3331 1992 .}
\REF\progress{Y. Meurice,  G. Ordaz and V. G. J. Rodgers, work in progress.}
\REF\dis{There is a large amount of literature on this subject. References can
be found in {\it Phase Transitions, Cargese 1980}, M. Levy, J.C. Le
Guillou and J. Zinn-Justin Editors, (Plenum Press, New York, 1982). }
\REF\confl{Z. Roskies \PRB 24 5305 1981 ; J. Adler, M. Moshe and
V. Privman \PRB 26 3958 1982 ; G. Nickel and M. Dixon
\PRB 26 3965 1982 .}

\Pubnum={U.Iowa 94-15\cr
hep-lat/9409021}
\titlepage
\title{ The Elusive Asymptotic Behavior of the High-Temperature
Expansion of the Hierarchical Ising Model}
\author{Y. Meurice and G. Ordaz}
\address{Department of Physics and Astronomy, University of Iowa,
Iowa City, Iowa 52242, USA}
\vfil
\abstract
We present a differential formulation of the recursion formula of the
hierarchical model which provides a recursive method
of calculation for the high-temperature expansion.
We calculate the first 30 coefficients of the high temperature expansion
of the magnetic susceptibility of the Ising hierarchical model with 12
significant digits.
We study the departure from the approximation which consists of identifying
the coefficients with the values
they would take if a $[0,1]$ Pad\'e approximant
were exact. We show that, when the order in the high-temperature
expansion increases,
the departure from this approximation grows more slowly than for nearest
neighbor models. As a consequence, the value of the critical exponent
$\gamma $ estimated  using Pad\'e approximants  converges very slowly
and the estimations using 30 coefficients have errors larger than 0.05.
A (presumably much) larger
number of coefficients is necessary to obtain the critical exponents
with a precision comparable to the precision obtained for nearest neighbor
models
with less coefficients.
We also discuss the possibility of constructing models where a
$[0,1]$ Pad\'e approximant
would be exact.

\endpage

\chapter{Motivations, Main Results and Notations}

The hierarchical model\refmark\dyson
is a model for which
the renormalization group transformation\refmark\wilson
reduces to a simple recursion formula. This recursion formula
is a simple  integral equation which has been studied in great
detail.\refmark{\sinai }
More recently, interesting results concerning the analyticity and the location
of the complex zeroes
of the probability distribution associated with the
infrared fixed point
have been found.\refmark\koch
The hierarchical model
has a free parameter $\epsilon $
which is used in the $\epsilon-$expansion. By adjusting this parameter,
one can use the hierarchical model as
an approximation\refmark{\dyson - \gaw}
to nearest neighbor models in various dimensions.
In the case $D=3$, the hierarchical approximation yields values of the critical
exponents which agree within a few percent with the best estimates
for nearest neigbor models.

A difficult but interesting question is how to improve the hierarchical
approximation while keeping control on the complexity of the renormalization
group transformation. This question can be answered for gaussian models
using group-theoretical methods.\refmark\marseille
However, the extension of these methods to interacting models
is not straightforward because many more terms can be generated
perturbatively. The main problem consists in identifying the most
important perturbations which should be added to the hierarchical model
in order to obtain a critical behavior more similar to nearest neighbor models.
A possible indicator which could be used to achieve this goal is
to calculate the effect of a perturbation on the high-temperature coefficients.
Before evaluating the effects of these perturbations, it is necessary
to develop efficient methods
to calculate the high-temperature  expansion of the
(unperturbed) hierarchical model.
Recently, analytical formula for the first four coefficients of the magnetic
susceptibility
were calculated for a Ising measure.\refmark\high
The method used relies
on high-temperature graphs. Due to the non-locality of the model,
the terms appearing in the coefficients
rapidly proliferate and require elaborate symbolic methods to be handled.

In this article, we present a recursive method
which allows an efficient calculation of the high-temperature expansion
of the hierarchical model. This method combines the renormalization
group method whith the high-temperature expansion but does not
rely on any graphical analysis.
It is a simple differential version
of the recursion relation (with an arbitrary rescaling of the spin variable).
It essentially solves the problem of the high-temperature expansion
provided that
we keep the numerical stability and the volume effects under control.
This method is explained in section 2 where we also
calculate the first 30 coefficients of the high-temperature expansion
of the magnetic susceptibility of the hierarchical Ising model
in a large ($2^{60}$ sites) but finite volume.
Checking with analytical results and comparing different implementations
of the differential recursion formula, we claim that for $\epsilon=1$,
the method is numerically stable as far as 12
significant digits are concerned, and that within this precision,
the first 30 coefficients
have reached their infinite volume
limit when $2^{60}$ sites are reached.
We also briefly discuss the possible use of the differential formula to obtain
analytical results concerning the high-temperature expansion.

In section 3,
we study the departure from the approximation which consists of identifying
the coefficients with the values they would take if the $[0,1]$ Pad\'e
approximant for the logarithmic derivative of the susceptibility
were exact.  The resulting approximate relation among the coefficients
was suggested by a numerical study
\refmark\num where we
calculated (without any approximation) the magnetic susceptibility
of the hierarchical
Ising model with up to $2^{18}$ sites. In this study,
we found that the numerical data can be fitted very precisely
with a simple power law of
the form $(1- \beta /\beta _0 )^{- g}$
in the {\it whole} high-temperature region, i.e., for $\beta \in [0,\beta _c)$.
The values of $g$ obtained
from the numerical data or from the $[0,1]$
approximant, differ significantly from
the  best estimates of the values of the  critical exponent $\gamma $
calculated \refmark{\sinai,\epsi} with the $\epsilon$-expansion.
In the case of the numerical data,
the most plausible explanation provided by a
a renormalization group analysis\refmark\num is that this discrepancy
is due to the limited size of the lattice used in these calculations.
As we shall recall in the following section, if the lattice has $2^n$ sites,
the recursion formula can only be used $n$ times, and $n=17$ does not seem to
be enough to get rid of the irrelevant components of the measure.
On the other hand, the fact that a $[0,1]$ approximant gives poor estimate
is not surprising. What is surprising is how slowly
the departure from this approximation grows when the order in the
high-temperature expansion increases.

In section 4, we use Pad\'e approximants to estimate $\gamma $ and $\beta_c$.
We first discuss how the results depend on the volume.
We show that when the volume is not too large - less or much less
than $2^{20}$ sites
for the orders considered here - the approximants can resolve a pair
of complex conjugate roots approaching the real axis in the complex
temperature plane. However, for larger volume, two real roots appear
with - in a large majority of cases - only one root stabilizing
near $\beta _c$. This stabilization toward the infinite volume value
seems to occur with a
precision comparable to the precision of the coefficients.
We have checked this statement for the six
approximants which can be calculated from the exact\refmark\high
values of the first four coefficients at infinite volume.
We concluded that the errors on the estimation of the critical
quantities due to finite volume or round-off errors were very small
compared to the size of the effects discussed below. More precisely,
the 12 significant digits precision on the infinite volume coefficients
are more than sufficient to discuss effects of order $10^{-3}$ in the critical
quantities. The results for the 435 approximants that one can obtain from
the coefficients
up to order 30 are then presented. The estimates of $\beta _c$ are slightly
too high for low order approximants but seem to converge reasonably well (i.e.,
small
errors in the fourth significant digit in most of the cases) when the order
becomes large enough. On the other hand, the estimates of $\gamma $ are
significantly
too high for low order approximants and evolve very slowly toward our best
estimates.
In order to give an idea, absolute errors of order 0.07 and 0.11 (respectively)
are typical
of the large order estimations for $\epsilon=1$ and 0 (respectively).

These large discrepancies can be understood in terms of the approximate
relations discussed in section 3. For instance,
the values of the critical exponent
$\gamma $ extracted from a $[9,10]$ Pad\'e is still midway in
between the crude estimate given by the $[0,1]$ approximant
(e.g., 1.52 for $\epsilon =1$) and the exact
value (1.30, in the same case). This point is discussed
at length in section 4 for $\epsilon =1$ and 0.
{}From these results, it appears that a (presumably much) larger
number of coefficients is necessary to obtain the critical exponents
of the hierarchical model
with a precision comparable to these
obtained\refmark\gaunt for nearest neighbor model
(i.e., a precision of 0.001 or better).

We also discuss (section 5) the possibility of constructing models where a
$[0,1]$ Pad\'e approximant
would be exact.
We first compare the approximate relations,
for the hierarchical model and the nearest neighbor models when the dimensional
parameter $D=4-\epsilon $ is varied.
We show that the corrections to the Gaussian result (for which the
approximation
is exact) appear only at second order in the $1/D -$ expansion\refmark\fisher
for the
hierarchical model while it appears in first order
for nearest neighbor models. Unfortunately, the use of this
expansion is obscured by a proliferation of zeroes and poles
appearing in the expression of
the coefficients when $D$ is promoted to a complex variable.\refmark\high
We also mention the fact that supersymmetric models\refmark\veneziano
might be used as a guide
to achieve this goal.

In the conclusions, we give various estimates of
the number of coefficients necessary to calculate $\gamma $ with a reasonable
precision. We discuss briefly the feasibility of the project
and additional problems which could be considered in this context.
We also compare with other recursive methods\refmark\creutz used to
calculate the high-temperature expansion of the $3D$ Ising model.
This raises the following dilemma: on one hand, we have a much more simple and
efficient
method of calculation, on the other hand, we need much more coefficients to
obtain a
comparable information.

In order to give a self-contained presentation, we briefly remind our
conventions
concerning the hierarchical Ising model and its free parameter.
Hierarchical models\refmark\dyson
are specified by a non-local hamiltonian bilinear
in the spin variables and a local measure of integration.
In the following, we consider exclusively the case of an Ising measure, where
the spin variables
take only the values $\pm 1$.
The models considered here have $2^n$ sites. In the following, $n$ will always
be used in this sense.
We label the sites with $n$
indices $x_n ..... x_1$, each index
being 0 or 1 . The energy of a spin configuration reads
$$H=-{1 \over 2} \sum\limits_{l=1}^{n}({c \over 4})^l
\sum\limits_{x_n,...,x_{l+1}}\ (\ \sum\limits_{x_l,....,x_1} \sigma
_{(x_n,....,x_1)} )^2\ . \ \eqno(1.1) $$
The motivations for this construction and the derivation of the
recursion formula
are reviewed
in Ref. [\num].

The model has a free parameter $c$ for which we shall use the
parametrization
$$c=2^{1-{2\over D}}.\eqno(1.2)$$
The parameter of the epsilon-expansion can be defined as
$$\epsilon = 4-D \eqno(1.3).$$
When $D\geq 4$, the model has a trivial continuum limit.\refmark\sinai
When $D\leq 2$, the model does not have a phase transition at finite
temperature.\refmark\dyson
These two rigorous results can be understood heuristically
in terms of the the self-intersection properties of the random walk
associated with $H$, by noticing that the Haussdorff dimension
of this random walk is $2/D$.\refmark\rw
{}From the point of view discussed at the beginning of the introduction,
the most interesting
region is $2\leq D \leq 4$ (i.e., $1\leq c \leq 2^{1/2} $), however,
the model has a
sensible infinite volume limit for a wider range of the free parameter, namely
$0<D<\infty$
 (i.e., $0<c<2$).
We define the magnetic susceptibility per site as
$$\chi_{_n}(\beta)\ = \ {1\over {2^n}}<\ (\sum _x \sigma _x )^2 \ >_n
\eqno(1.4)$$
The high-temperature expansion of this quantity reads
$$\chi_n (\beta)=1\ + \ b_{1,n}\beta \ + \ b_{2,n}
\beta ^2 \ + \ ... \eqno(1.5)$$

\chapter{A Recursive Method for the Calculation of the High-Temperature
Expansion}

Due to the non-locality of the hierarchical model, the evaluation
of the coefficients is rather tedious.
We have used three types of techniques to calculate these
coefficients. Two of them have been described at length in Ref. [\high].
Up to order 4, analytical calculations can be performed using
algebraic methods\refmark\high and yield
expressions where $D$ and $n$
are arbitrary. Substituting numerical values into these formulas
provides very precise numerical comparisons with other numerical methods.
We checked the validity of the analytical results by
introducing a high-temperature expansion in the numerical method used
in Ref. [\num].
This numerical method is very reliable but rather slow
and does not allow large $n$ calculations, because the computer time
involved grows like $4^n$.

In order to perform large $n$ calculations, we have designed another
method of calculation based on the Fourier transform version of the
recursion relation.
If we call $\widehat{P}_n(k)$ the Fourier transform of the probability
of the mean spin for $2^n$ sites, we obtain the recursion relation
$$\widehat{P}_{n+1}(k)=C_{n+1}\exp(-{1\over 2}\beta
({c\over 4})^{n+1}{{\partial ^2} \over {\partial k ^2}})(\widehat{P}_n(k))^2
\ , \eqno(2.1)$$
where $C_{n+1}$ is adjusted in order to get $\widehat{P}_{n+1}(0)=1$
and the exponential of the second derivative is defined through
the Taylor expansion.
For the Ising model considered here, the initial function is
$$\widehat{P}_{0}(k)=\cos(k)\ .\eqno(2.2)$$
In this formulation, the magnetic susceptibility reads
$$\chi_n (\beta)=-{1\over {2^n}}
{{\partial ^2} \over {\partial k ^2}}\widehat{P}_n(k) \bigg| _{k=0}
\ .\eqno(2.3)$$
More generally, $\widehat{P}_n(k)$ generates the average values for
the even powers of the mean spin:
$$\widehat{P}_n(k)=\sum_{m=0}^{\infty}{{(-ik)^{2m}}\over{2m!}}
<\ (\sum _x \sigma _x )^{2m}\ >_n \eqno(2.4) $$

Equation (2.1) is a differential formulation of the recursion
formula. Unlike the original formulation, it does not involve any
integrals and it can be reduced to a sequence of purely
algebraic operations when expanded in powers of $\beta $ to a given
order.
In other words, (2.1) can be seen as a recursive definition of the
high-temperature
expansion of the average values of the even powers of the mean spin.
This simple
procedure completely solves the problem of the high-temperature expansion
of the hierarchical model, provided that the operations can be performed
within a reasonable amount of time and provided that we can reach the infinite
volume limit with an acceptable numerical precision.

The computer time necessary to calculate the
$m$ first coefficients grows like $m^2 n$
since we only need to retain the first $2m+2$ terms in the Taylor
expansion of the functions introduced in the recursion formula.
This is much better then the $4^n$ behavior reached by straightforward
evaluation\refmark\high following the integration method of Ref. [\num ].
In order to give an idea concerning the absolute time scale, it takes
a little less than two days, with Mathematica,
to iterate 60 times Eq. (2.1) expanded up to order 30.

Since the method requires many
iterations, it is important to check the numerical stability of the algorithm.
One potential numerical difficulty is the appearance of
large numbers in (2.4), since one expects (see section 3 of Ref.[\num]),
that for large $n$,
$$<\ (\sum _x \sigma _x )^{2m}\ >_n \ \propto 2^{nm}\ . \eqno(2.5)$$
This difficulty
can be overcome either by replacing (2.1) by a recursion formula for the
logarithm of $\widehat{P}_n$
divided by the number of sites, or
by introducing an appropriate rescaling in (2.1). In the following, we only
discuss the magnetic
susceptibility and consequently, the second possibility is the most convenient
one.
In general, a rescaled version of (2.1) can be obtained by defining
$$R_{\lambda ,n}(k)\ = \ \widehat{P} _n ({k\over{\lambda ^n}}) \ . \eqno(2.6)$$
In terms of this rescaled quantity, (2.1) becomes
$$R_{\lambda ,n+1}(k)=C_{n+1}\exp(-{1\over 2}\beta
({c\over 4}\lambda ^2)^{n+1 }{{\partial ^2} \over {\partial k ^2}})(R_{\lambda
,n}({k\over \lambda}))^2
\ , \eqno(2.7)\ .$$
The large volume behavior shown in (2.5), can then be compensated by
chosing $\lambda = \sqrt{2}$. This choice has been used for all the
numerical results given hereafter. This modification
improves significantly the precision of the calculation as discussed below.
Note also that
the renormalization group transformation is obtained by taking $\lambda =
2c^{-1/2}$
in (2.7).

In the rest of this section and the next section, we discuss mostly
the numerical results for $D=3$. Results for other values
of $D$ will be discussed in sections 4 and 5.
Checking the numerical results obtained with the rescaled
recursion formula (2.7) at $\lambda=\sqrt{2}$, for the first four coefficients,
for all $n$ up to 60, with the exact results at finite volume,\refmark\high
we found errors of at most 1 in
the 13-th significant digit. This precision is almost twice better
than the one obtained with the original formula (2.1).
We have also compared slightly different
numerical implementations for which the round-off errors should be
different and found agreement at that level of precision for the
30 coefficients. We conclude that after 60 iterations, the numerical
errors are at most in the 13-th significant digit.

The next question is to determine how close the coefficients calculated
with $2^{60}$ sites are from the infinite volume ones.
In order to get a first
idea concerning the volume dependence of the coefficients,
the results for $D=3$ and $n=10,\ 20,\ 30$ and 40 are given in Table 1.
It is possible to compare the values of the first
four coefficients with their infinite volume limit values:\refmark\high
$$\eqalign{b_1=&1.242602432206561\cr
b_2=&1.280296258947416\cr
b_3=&1.249766644307651\cr
b_4=&1.171252787942138\ .}\eqno(2.8) $$
One sees that roughly, the agreement with the infinite volume
limit improves by two digits each time that $n$ is increased by 10.
This suggests that
the volume dependence is exponentially suppressed.
In order to check this possibility, we have plotted in Fig.1 the logarithm
of $b_{m,60} - b_{m,n}$ for $n<60$ and $m=$1, 5, 10, 15, 20, 25, and 30.
We have selected these few values of $m$ in order to keep the distinction
among various values of $m$ visible on the figure.
The trajectories for other $m$ fall approximately
within the envelope delimited by
the ones displayed.
The figure makes clear the length of the plateau at low $n$ increases with $m$,
in other words, higher
order coefficients take a larger volume to stabilize (with exponential
precision)
near their infinite volume
limit. After this, all the trajectories
appear more or less parallel. The slope can
be compared with the
analytical results\refmark\high which show that the leading $n$-dependence
of the first four coefficients comes from terms
proportional to $({c \over 2})^n$.
We remind that in
the case considered here, ($D=3$),
 $c=2^{1/3} $.
The general agreement with this prediction is quite clear in Fig.1.
In conclusion, Fig. 1 supports convincingly the
possibility that the coefficients displayed reach their infinite volume limit
with an exponential precision, and that up to the 30-th coefficient for $D=3$,
the $n=60$ results give estimates of the infinite volume values
which are accurate up to the 12-th significant digits (which is the
limit of our numerical precision). The numerical results are given in Table 2
for $ n=60$ and $D=3$ and also, for further reference, for $D=4$.

For comparison, we have also given in Table 1 the first 17 coefficients
for the nearest
neighbor model in three dimensions on a simple cubic lattice. These
coefficients have been calculated from the tables given in Ref. [\gaunt ].
Note that these tables are given for the expansion parameter $tanh(\beta )$,
and an additional expansion is necessary
in order to obtain an expansion of the form of
Eq. (2.5).
In addition, we have rescaled the temperature of the nearest neighbor model
in such a way that its first coefficient
coincides with the one of the hierarchical model
in the infinite volume limit. The large discrepancies indicate that
the high-temperature coefficients should be sensitive to
``non-hierarchical'' perturbations.

Before closing this section, let us mention the possibility of using Eq. (2.7)
to obtain analytical results. This equation relates the average value of
the $2m$-th power of the mean spin expanded to a certain order in $\beta$
to average values of larger $m$ and lower order in $\beta $. This suggest
a bootstrap procedure which is presently under investigation.
Note also that the $(c/2)^n$ rule for the volume dependence
appearing in Fig. 1, can be inferred from
Eq. (2.7) with $\lambda=\sqrt{2}$ together with (2.5).
Finally, let us mention that when $n$ goes to infinity, the recursion
formula has a fixed point of the form $exp(-{1\over 2}\chi k^2)$.
The linearized transformation about this fixed point is presently under study.

\chapter{Approximate Relations among the High-Temperature Coefficients}

In a recent numerical calculation,\refmark\num
we found that the numerical data for the susceptibility
at $n=14$, 15 and 16 can be fitted very precisely
with a simple power law of
the form $(1- \beta /\beta _0 )^{- g}$
in the {\it whole} high-temperature region.
As a consequence, it is possible to obtain a simple approximate formula for
the high-temperature coefficients of the susceptibility
in terms of $g$ and
$\beta _0$ by comparing Eq. (2.5) and
\def\bo{\beta _0 }
$$(1- \beta / \beta _0 )^{- g} = 1+{g\over \bo}
+{{g(g+1)}\over {2 \bo ^2}}+....\eqno(3.1)$$
Approximate values for $g$ and $\bo $ are obtained by comparing the two
first terms of Eqs. (3.1) and (2.5). Solving
${g\over \bo}=b_{1,n} $ together with  ${{g(g+1)}\over {2 \bo ^2}}=b_{2,n}$, we
obtain
$$\eqalign{g \ &= \ ({{2b_{2,n}}\over{b_{1,n}^2}}-1)^{-1}\cr
\bo \ &= \ ({{2b_{2,n}}\over{b_{1,n}}}-b_{1,n})^{-1} \ .}\eqno(3.2)$$

By comparing Eq. (2.5) with Eq. (3.1), and plugging the values of $\bo $ and
$g$ given by
Eq. (3.2), we obtain approximate values of the coefficients denoted
$b^{appr.}_{m,n}$.
As an example, at third order we obtain
$$b^{appr.}_{3,n}=
{{g(g+1)(g+2)}\over {6 \bo ^3}}\ .\eqno(3.3)$$
Using Eq. (3.2), we then obtain
$$b^{appr.}_{3,n}={b_{2,n} \over 3}\big( {{4 b_{2,n}}
\over{b_{1,n}}}-b_{1,n}\big) \ .
\eqno(3.4) $$
Note that the approximate relations are homogeneous in the sense that
at order $m$, all the terms have the form $b_{1,n}^q b_{2,n}^p$ with
$2p+q=m$.

In order to see how well these approximate relations are satisfied, we have
calculated the difference between the exact and the approximate value of the
coefficients
denoted
$$\Delta _{m,n} = b_{m,n}-b^{appr.}_{m,n}\ . \eqno(3.5)$$
Note that due to the homogeneity of the
approximate relations mentioned above,
the ratio ${{\Delta _{m,n}}/{b_{m,n}}}$
is invariant under a rescaling of the temperature.
In particular, this implies that the rescaling
of $\beta $ used in Table 1 for the nearest neighbor model has no effect
on this ratio.
The numerical values of this ratio for
$D$=3, $m$ up to 30, and $n=$ 40 and 60
are displayed in Table 3. The comparison
between the results for the two values of $n$, shows
that the volume dependence
is not an issue for the discussion in the rest of this section.

We can now
compare the approximate relations for the hierarchical model and the nearest
neighbor model. We see that in
both cases the quality of the approximation deteriorates
when $m$ increases. However, the rate are quite
different: in the nearest neighbor
case, $\Delta _{m,n}/{b_{m,n}}$ increases by approximately 0.04 each
time $m$ is increased by
1, while the corresponding increase is approximately 0.02 (and decreases with
$m$)
for the
hierarchical model.

Equation (3.2), is exactly what we would have obtained from a $[0,1]$
Pad\'e approximant
for the logarithmic derivative of the susceptibility.
In the infinite volume limit, we obtain
$g=1.5189$ and $\bo =1.2224 $.
The value of $g$ is
significantly larger than
the value of the critical exponent $\gamma $
obtained\refmark\epsi in the $\epsilon$-expansion, namely 1.300.
A value close to 1.30 has also been
obtained with independent methods by Bleher.\refmark\sinai
The value of $\bo $ is also significantly larger than our best
estimate\refmark\num
$\beta _c = 1.179$.
It is not
a surprise that a $[0,1]$ Pad\'e approximant
provides bad estimates of the critical quantities.
What is a surprise is how slowly the high-temperature series of the
hierarchical model
departs
from this poor approximation when the order $m$ increases.
This slower rate of departure implies
that a number of coefficients much larger than in the nearest neighbor case,
is necessary in order to
estimate the critical quantities with a comparable precision.

We suspect that there exists a definite relationship between the quality of the
estimates made from finite $n$ and infinite
$m$ (as done in Ref.[\num]) and those made from finite $m$ and infinite (or
sufficiently large) $n$
(as done here). This remark is motivated by the observation that the values of
$\gamma $ obtained
at $n=16$ (e.g., 1.47 for $D$=3)
are close to those obtained from low order Pad\'e approximants (see next
section).
The discrepancy of the finite
volume fits with the $\epsilon-$expansion was attributed\refmark\num
to the fact that for $n=16$, for instance,
we can only use the recursion formula 16 times, while a larger
number seems necessary to get rid of the irrelevant components of
the measure. This statement has a counterpart for the
high-temperature expansion which is roughly
the following: the coefficients which
at a finite $n$, have reached
their infinite value with an acceptable precision, provide estimates of
the critical exponents consistent with the finite $n$ data.
Since a direct calculation of the susceptibility at $n=30$ or 40 seems excluded
with the exact method used in Ref. [\num ],
we cannot sharpen further the statement made above. However, the $n-$dependence
of the numerical estimates shown in Figs. 11 and 12 of ref. [\num], the
regularity of the length of the shoulders of Fig. 1 of this article
and the results shown below suggest that a more quantitative study could
be made in the future, possibly using the Eq. (2.7).

\chapter{Estimation of $\beta _c$ and $\gamma$ Using Pad\'e Approximants}

In this section, we discuss the estimation of $\beta _c$ and $\gamma$, using
$[L,M]$ Pad\'e approximants with $L+M+1\leq 30$. In the following, we call
$L+M+1$ the $order$ of the approximant.
Since our calculation of the coefficients has been done
at large but finite volume, we will first address the question of the volume
dependence of the results obtained. As well-known, at finite volume, a
singularity of the susceptibility on the real
temperature axis is impossible. However, when the volume
increases, a pair of complex conjugate singularities can approach arbitrarily
closely the real axis. As we now proceed to explain,
the Pad\'e approximants sense this approach of the real axis,
but with a finite resolution.

We have studied a large set
of approximants using the finite volume
coefficients and followed the motion of the roots of the denominator
when the volume is increased. We have observed the following patterns in
a large majority of cases.
For $n$ below a number which is usually
between 10 and 15, the roots move by large steps,
usually of the same order as the change in the coefficients. During this
preliminary
motion, a pair of roots ``pinches'' the real axis and two clearly distinct real
roots appear. This situation is illustrated in Fig. 3. One of these two  real
roots
almost immediately stabilizes near $\beta _c$ (1.179 in this case), while the
other makes a few erratic steps
before stabilizing (more slowly) near another real value clearly distinct from
$\beta_c$. Fig. 4 illustrates the small size motion of all the roots
in the complex plane for $n$
between 17 and 60. All the roots are well separated and for the roots labeled
1, 3, 4, 5, and 6 the changes
are barely perceptible. We have skipped $n=$14, 15 and 16, because the root
labeled as 2 in Fig. 3
makes sudden changes during these 3 steps.
Comparing with other cases, it appeared clearly that the value of $n$
for which the pinching of the real axis occurs increases with the order
($L+M+1$)
of the approximant.
In other words, it appears that the Pad\'e approximants of a given order
can only resolve a pair of complex conjugate roots when their distance
to the real axis is not less than a certain value (which
decreases with the order of the approximant).

In most of the cases, the roots are widely
separated for $n>20$. However, there are a few exceptional cases
where two real roots are very close to $\beta_c$, for instance the $[13,12]$
and the $[12,13]$ approximants for $D=3$, and also $[10,11]$ and $[11,12]$
for $D=4$.
In these special cases, the residues
at both poles are very different from the residues at the pole near $\beta _c$
for most approximants of comparable order.

A closer look at the numbers indicates that the change in the
location of the pole near 1.18 and the change (due to the volume)
in the value of the residue at
the
pole
decrease exponentially when the volume
is increased. In the case of approximants requiring 4 coefficients or less,
we can indeed obtain a precise comparison between the results at $n=60
$ and $n=\infty $ using the infinite volume limit values of Eq. (2.8).
We found that for the six approximants with $L+M+1\leq4$ the difference
between the $n=60$ and $n=\infty$
appears at worse in the 11-th
significant digit for both $\beta _c$ and $\gamma $.
We also made independent random changes of order $10^{-4}$ in the values of a
few coefficients
and did not observe any instability in the estimations.
Our general conclusion concerning the volume dependence of the estimations
obtained
from Pad\'e approximants calculated with coefficients corresponding to models
with $2^{60}$
sites, is that the results obtained approximate the infinite volume results
with a precision which is far beyond what would be required to
discuss effects of the order of $10^{-3}$ in the critical quantities.

We can now present the numerical results.
In Table 4 and 5, we give the values of $\beta _c$ and $\gamma$
obtained from the $[j+k,j]$ approximants for $1\leq j \leq 14$, $k=\pm 1,0$
and $D=3$ and 4.
In both cases, we have used the coefficients for $n=60$.
We can now compare these results with our best estimates.
In the case $D=3$, one observes a very slow (compared to nearest neighbor
models) progression toward
$\beta _c =1.179$ and\refmark{\sinai, \epsi} $\gamma =1.300 $
when the order of the approximants is increased.
In the case $D=4$, a similar progression toward
$\beta _c =0.665 $
and the trivial value $\gamma=1$.

In order to give a more complete idea concerning
the results obtained
from Pad\'e approximants, we give in Fig.5, the distribution of
of values $\gamma $
obtained from the 435 different approximants of order less than 30
for $D=3$.
Less than one half of one percent of the data fell outside of the figure.
The figure shows separately the distribution for $L+M\leq 10$, $10<L+M \leq20$
and $20 <L+M <30$. The figure makes clear that the distribution
gets more peaked and that the average decreases when the order of the
approximants
increases.
The average for the distribution with $20<L+M<30$ is 1.382
which is still far away from 1.300. Note also that the number of approximants
giving a value of $\gamma $ near 1.30 is non-negligible for $L+M<10$,
however, this feature clearly fades away when the order increases.

\chapter{Are There Models for which the Approximate Relations are Exact ?}

Up to now, the approximate relations discussed in section 3 have had
a rather unpleasant effect: we need to calculate many coefficients
in order to get any reliable result. However, the situation would be
opposite if these relations were exact or if the departure could
be estimated very precisely.
It is thus tempting to try to construct
models where, for instance, the $[0,1]$ Pad\'e approximant would be exact.

A modest step in this direction consists in
considering the $D-$dependence of the approximate relations
studied in the special case $D=3$ in section 3,
and compare with the nearest neighbor models.
In the rest of this section, we consider the infinite volume limit
and the reference to $n$ disappears.
In Fig. 2, we have plotted, $\Delta _m /b_m $ for $m=3$ and 4 as a function
of $D$ using the exact analytical results of Ref. [\high].
We see that both quantities decrease when $D$ becomes large and vanish for
special values
of $D$ between 1 and 2. A more detailed graph would show that $\Delta _3$
vanishes
near $D=1.14$ and $\Delta _4$ near $D=1.66$.
It is not clear that we can learn anything
interesting from the behavior of the coefficients in this low $D$ region.
On the other hand, the fact that the approximation becomes better for large $D$
is due to the fact that, as for the nearest neighbor models\refmark\fisher,
the corrections to the Gaussian result $b_m/b_1^m =1$ vanish.
Note that the approximation discussed in section 3 is more general
than the Gaussian approximation, since it allows $b_2$ to be distinct
from $b_1^2$, however it is obviously exact in the gaussian approximation.

We first consider the nearest neighbor case.
After an elementary calculation and an appropriate expansion of $tanh(\beta)$,
we obtain that for the Ising models
with nearest neighbor interactions on an hypercubic lattice of dimension $D$:
$b_1=2\,{D}$,
$b_2=-2\,{ D} + 4\,{{{ D}}^2}$,
$b_3={{4\,{ D}}\over 3} - 8\,{{{D}}^2} + 8\,{{{D}}^3}$,
$b_4={{10\,{ D}}\over 3} + {{16\,{{{ D}}^2}}\over 3} - 24\,{{{D}}^3} +
  16\,{{{ D}}^4}$.

Expanding $\Delta_m /b_m$ in $1/D$, we obtain:
$$(\Delta_3 / b_3 ) ={1\over {6\,{D}}} - {1\over {36\,{{{ D}}^3}}} -
  {1\over {36\,{{{ D}}^4}}} - {5\over {216\,{{{D}}^5}}} +
  {{{\rm O}({1\over {{ D}}})}^6}\eqno(5.1)$$
and
$$(\Delta _4 /b_4 )= {5\over {12\,{ D}}} - {1\over {4\,{{{D}}^2}}} -
  {1\over {18\,{{{ D}}^3}}} - {{25}\over {288\,{{{ D}}^4}}} -
  {{103}\over {1728\,{{{ D}}^5}}} + {{{\rm O}({1\over {{D}}})}^6}
\eqno(5.2)$$

We can now compare with a similar expansion for the hierarchical model.
Using the results of section 6 of Ref. [\high ], we obtain
$$(\Delta_3 / b_3 ) =
{{8\,{{\log (2)}^2}}\over {3\,{{{D}}^2}}} -
  {{512\,{{\log (2)}^3}}\over {9\,{{{D}}^3}}} +
  {{2488\,{{\log (2)}^4}}\over {3\,{{{ D}}^4}}} -
  {{266432\,{{\log (2)}^5}}\over {27\,{{{ D}}^5}}} +
    {{{O}({1\over {{D}}})}^6}\eqno(5.3)$$
and
$$(\Delta_4 / b_4 ) ={{20\,{{\log (2)}^2}}\over {3\,{{{D}}^2}}} -
  {{1096\,{{\log (2)}^3}}\over {9\,{{{D}}^3}}} +
  {{80084\,{{\log (2)}^4}}\over {63\,{{{D}}^4}}} -
  {{2764360\,{{\log (2)}^5}}\over {441\,{{{D}}^5}}} +
  {{{O}({1\over {{\rm D}}})}^6}\eqno(5.4)
$$

It is clear that the corrections to the Gaussian result appear
at an higher order ($(1/D)^2$) for the hierarchical model than for the
nearest neighbor model. However a direct comparison of the
two expansions does not shed any new
light on Table 3. The reason being that for $D=3$, the
series given in Eqs. (5.3-4) diverge. The origin of these divergences
can be found from the zeroes and
poles of the coefficients in the complex $1/D -$plane (see Ref. [\high]
for more detail). For instance, in the case of Eq. (5.3), we found
a pair of poles at $(1/D)=-0.15608\pm i  0.042037$ and then
at $(1/D)=0.25$ etc.... A complete description seems to be of little
interest, since it does not suggest any way to improve the approximation.

We shall end this section with a speculative note.
Supersymmetric gauge theories provide non-trivial
examples where the dependence of some
Green's functions on the mass is expected to be a
simple power law,\refmark\veneziano
which is somehow the same thing as saying that Eq. (3.1)
is exact. It is clearly difficult
to compare these supersymmetric
theories, which as far as we know do not have a satisfactory
lattice regularization, with
the spin models considered here.
However, the non-locality of the hierarchical model suggests
that it is an ``effective'' theory
obtained  after integrating over some other local variables.
Indeed, a concrete example has been given in Ref. [\conje], where
the hierarchical model was reformulated as a local theory with
additional spin variables integrated with a Gaussian measure.
It thus conceivable that one could find a reformulation of
the hierarchical model where approximate symmetries
playing a role similar to the supersymmetries, could be
more manifest.

\chapter{Conclusions}

We have presented a recursive method of calculation which allows an
efficient calculation of the high-temperature expansion of the hierarchical
Ising model.
The number of coefficients calculated
is larger than what seems achievable\refmark{\gaunt\creutz} in the case of
nearest
neighbor models.
However, our errors on
the critical exponent $\gamma $ range between $0.05$ and $0.1$
and are much larger than the typical precision (0.001)
obtained for nearest neighbor models with less than 20 coefficients.
We have carefully considered the errors due to the fact that we used a large
but finite volume to calculate the coefficients and concluded that these errors
were much smaller than the effects described above.

We have identified the reason of this slow approach of the asymptotic regime
by comparing the departure from the approximation corresponding to a $[0,1]$
Pad\'e approximant. The last two columns of Table 3 indicate that at least
50 coefficients will be necessary in order to get a precision of 0.001
for $\gamma $. However, Tables 4 and 5 seem to leave open the possibility
that a much larger number might be necessary.
With the numerical method used in this paper, it would probably take
a year to calculate 400 coefficients.
We are presently working
on an optimized version\refmark\progress
of the existing program which would allow us
to achieve this goal in a shorter time.

Note that recursive methods
have been found\refmark\creutz to calculate the high temperature
expansion of the $3D$ Ising model with nearest neighbor interactions. Their
recursive
step consists of adding one spin and the calculation requires
large parallel computers
to calculate 24 coefficients. Our recursive step
consists of putting in contact two identical systems of size $2^n$ in order to
get a system
of size $2^{n+1}$ and 24 coefficients can be obtained within less than two days
with a PC.
On the other hand, with 24 coefficients, we obtain estimates which are less
precise by one
order of magnitude. We would like to check weather or not the following
possibility occurs:
our recursive algorithm would require a logarithmically smaller time of
computation,
but at the same time require an exponentially larger number of coefficients
in order to reach a given precision.

Having at hand a large number of coefficients for the high-temperature
expansion would allow precise comparison with the $\epsilon-$expansion
for which a large number of coefficients is available.\refmark\epsi
In the case of nearest neighbor models, discrepancies have been observed
between these two methods.\refmark\dis
However, several authors\refmark\confl
have shown that these discrepancies can be removed
by an appropriate treatment of the confluent singularities.
For the hierarchical model, this is a completely open question.
It is also possible that the simplicity of Eqs. (2.6-8) would allow
to obtain analytical results concerning the asymptotic behavior of
the high-temperature expansion.

\ack
One of us (Y.M.) would to thank the
the theory group of
Brookhaven National Laboratory
and the Institut de Physique Th\'eorique at Louvain-la-Neuve
for their hospitality while
part of this work
was completed.

\refout
\vfill
\eject
\centerline{FIGURE CAPTIONS}

\noindent
Fig.1: The logarithm of the difference between
the largest volume value $b_{m,60}$ and $b_{m,n}$ for $n<60$ is
plotted for $m=$1, 5, 10, 15, 20, 25 and 30. The solid line is the the
analytical slope for low $m$. The value of $D$ is 3.

\noindent
Fig.2: The difference between the exact and approximate values
of the third and fourth coefficient divided by the exact value,
is shown as a function of $D$. The analytical results of Ref. [\high]
in the infinite volume limit were used to draw the curve.

\noindent
Fig.3: The motion of the pair of complex conjugate roots near the
critical value for $n$ between 6 and 13. A $[8,8]$ Pad\'e approximant
was used for $D=3$.

\noindent
Fig.4: The motion of all the roots
for $n$ between 17 and 60 with a $[8,8]$ Pad\'e approximant
and $D=3$.

\noindent
Fig.5
Distribution of the values of $\gamma $ obtained from Pad\'e approximants
with $L+M+1\leq 30$ and $D=3$.

\centerline{TABLE CAPTIONS}

\noindent
Table 1: The coefficients of the high-temperature expansion of the
susceptibility up to order 20 for the hierarchical model with $D=3$
and $2^n$ sites with $n=$10, 20, 30 and 40. The last column is for the
nearest neighbor model on a simple cubic lattice (see Ref. [\gaunt ]).
The temperature of the nearest neighbor model
has been rescaled in such way that the first coefficients of the two
last column coincide.

\noindent
Table 2: The coefficients of the high-temperature expansion of the
susceptibility up to order 30 for the hierarchical model with $D=3$
and 4
with $2^{60}$ sites.

\noindent
Table 3: The difference between the exact and approximate values
of the first 30 coefficients divided by the exact value,
for the hierarchical model with $D=3$
and $2^n$ sites with $n=40$ and 60. The last column is
the same quantity for the
nearest neighbor model on a simple cubic lattice.

\noindent
Table 4: The values of $\beta _c $ and $-\gamma $ (between parenthesis)
for $D=3$
obtained from $[j+k,j]$ Pad\'e approximants, for $1\leq j\leq 14$ and $k=
\pm 1$ and 0. A star indicates two very nearby poles.

\noindent
Table 5:
Same quantities as in Table 4 but for $D=4$.
\vfill
\eject
\nopagenumbers
\def\med1{height.3cm depth.2cm}
\def\hrp{height.6cm depth.4cm}
\centerline{\vbox{\offinterlineskip
\halign{\vrule#&\quad#\hfil\quad&
\vrule#&\quad\hfil#\hfil\quad&
\vrule#&\quad\hfil#\hfil\quad&
\vrule#&\quad\hfil#\hfil\quad&
\vrule#&\quad\hfil#\hfil\quad&
\vrule#&\quad\hfil#\hfil\quad&\vrule#\cr
\noalign{\hrule}
\hrp&\multispan9 \hfil $b_{m,n}$ {\bf
  Hierarchical Model} $D=3$\hfil&&
  $b^{sc}_m({b^{HM}_1 \over b^{sc}_1})^m$&\cr
&\multispan9\hrulefill&&Simple&\cr
height.3cm depth.2cm&m$\backslash$n&&10&&20&&30&&40&&Cubic&\cr
\noalign{\hrule}
\med1&1&&1.2258497&&1.2424374&&1.2426008&&1.2426024&&1.2426024&\cr
\med1&2&&1.2389539&&1.2798863&&1.2802922&&1.2802962&&1.2867173&\cr
\med1&3&&1.1820851&&1.2490897&&1.2497599&&1.2497665&&1.3146331&\cr
\med1&4&&1.0785253&&1.1703159&&1.1712435&&1.1712527&&1.2987587&\cr
\med1&5&&0.9628730&&1.0760627&&1.0772198&&1.0772312&&1.2804051&\cr
\med1&6&&0.8452212&&0.9755497&&0.9768990&&0.9769123&&1.2429628&\cr
\med1&7&&0.7316955&&0.8746523&&0.8761526&&0.8761674&&1.2051755&\cr
\med1&8&&0.6274001&&0.7785983&&0.7802084&&0.7802243&&1.1594697&\cr
\med1&9&&0.5339023&&0.6893634&&0.6910447&&0.6910612&&1.1146199&\cr
\med1&10&&0.4510918&&0.6073781&&0.6090960&&0.6091130&&1.0663937&\cr
\med1&11&&0.3787313&&0.5329771&&0.5347017&&0.5347187&&1.0196561&\cr
\med1&12&&0.3163770&&0.4662812&&0.4679872&&0.4680040&&0.9718193&\cr
\med1&13&&0.2631594&&0.4069559&&0.4086227&&0.4086391&&0.9258158&\cr
\med1&14&&0.2179946&&0.3543998&&0.3560112&&0.3560270&&0.8799407&\cr
\med1&15&&0.1798494&&0.3079961&&0.3095398&&0.3095551&&0.8360504&\cr
\med1&16&&0.1478169&&0.2671874&&0.2686545&&0.2686690&&0.7929596&\cr
\med1&17&&0.1210802&&0.2314434&&0.2328281&&0.2328417&&0.7518815&\cr
\med1&18&&0.0988810&&0.2002368&&0.2015355&&0.2015483&&&\cr
\med1&19&&0.0805232&&0.1730514&&0.1742628&&0.1742747&&&\cr
\med1&20&&0.0653886&&0.1494050&&0.1505292&&0.1505403&&&\cr
}\hrule}}
\vskip 1.5truecm
\centerline{\bf Table 1}
\vfill\break
\end

\centerline{\vbox{\offinterlineskip
\halign{\vrule#&\quad#\hfil\quad&
\vrule#&\quad\hfil#\hfil\quad&
\vrule#&\quad\hfil#\hfil\quad&
\vrule#&\quad\hfil#\hfil\quad&
\vrule#&\quad\hfil#\hfil\quad&
\vrule#&\quad\hfil#\hfil\quad&\vrule#\cr
\noalign{\hrule}
\hrp&\multispan9 \hfil$\Delta_{m,n}\over b_{m,n}$ {\bf
  Hierarchical Model} $D=3$\hfil&&
 $\Delta^{sc}_m/b^{sc}_m$&\cr
&\multispan9\hrulefill&&Simple&\cr
\med1&m$\backslash$n&&10&&20&&30&&40&&Cubic&\cr
\noalign{\hrule}
\med1&3&&0.0158572&&0.0169702&&0.0169804&&0.0169806&&0.0540540&\cr
\med1&4&&0.0258695&&0.0305484&&0.0305913&&0.0305920&&0.1076487&\cr
\med1&5&&0.0380619&&0.0481560&&0.0482477&&0.0482490&&0.1751963&\cr
\med1&6&&0.0496359&&0.0670346&&0.0671910&&0.0671931&&0.2374947&\cr
\med1&7&&0.0591659&&0.0858405&&0.0860773&&0.0860803&&0.3020008&\cr
\med1&8&&0.0680981&&0.1056202&&0.1059488&&0.1059529&&0.3614195&\cr
\med1&9&&0.0764565&&0.1261950&&0.1266243&&0.1266296&&0.4191415&\cr
\med1&10&&0.0833798&&0.1467902&&0.1473290&&0.1473355&&0.4719078&\cr
\med1&11&&0.0887709&&0.1672670&&0.1679229&&0.1679306&&0.5216787&\cr
\med1&12&&0.0931015&&0.1878589&&0.1886366&&0.1886456&&0.5669304&\cr
\med1&13&&0.0965458&&0.2085752&&0.2094777&&0.2094881&&0.6089341&\cr
\med1&14&&0.0988445&&0.2291884&&0.2302178&&0.2302296&&0.6469778&\cr
\med1&15&&0.0997181&&0.2495106&&0.2506692&&0.2506823&&0.6819434&\cr
\med1&16&&0.0991350&&0.2695120&&0.2708010&&0.2708156&&0.7135224&\cr
\med1&17&&0.0972431&&0.2892454&&0.2906646&&0.2906805&&0.7423546&\cr
\med1&18&&0.0941709&&0.3087412&&0.3102893&&0.3103066&&&\cr
\med1&19&&0.0899061&&0.3279709&&0.3296459&&0.3296645&&&\cr
\med1&20&&0.0843037&&0.3468710&&0.3486704&&0.3486904&&&\cr
}\hrule}}
\vskip 1.5truecm
\centerline{\bf Table 2}

\vfill\break

\centerline{$D=3, n=40$}
\centerline{\vbox{\offinterlineskip
\hrule
\halign{\vrule height.3cm depth.2cm\quad#\hfil\quad&
\vrule height.3cm depth.2cm\quad\hfil#\hfil\quad&
\vrule height.3cm depth.2cm\quad\hfil#\hfil\quad&
\vrule height.3cm depth.2cm\quad\hfil#\hfil\quad\vrule height.3cm depth.2cm\cr
j$\backslash$k&-1&0&1\cr
\noalign{\hrule}
1&1.22239(-1.51895)&1.13546(-1.31059)&1.24663(-1.73445)\cr
2&1.18312(-1.44929)&1.19544(-1.49493)&1.18145(-1.42184)\cr
3&1.18963(-1.46908)&1.18785(-1.45988)&1.18799(-1.46077)\cr
4&1.18798(-1.46068)&1.18786(-1.45990)&1.18322(-1.42373)\cr
5&1.18963(-1.46792)&1.18393(-1.43066)&1.18360(-1.42726)\cr
6&1.18323(-1.42285)&1.18445(-1.43487)&1.18059(-1.38893)\cr
7&1.19340(-1.39791)&1.18185(-1.40813)&1.18153(-1.40335)\cr
8&1.18141(-1.40131)&1.18071(-1.38726)&1.18184(-1.40739)\cr
9&1.18162(-1.40439)&1.18265(-1.41495)&1.18098(-1.39523)\cr
}\hrule}}

\vskip 1.5truecm
\centerline{\bf Table 3}
\vfill\break

\centerline{$D=4, n=40$}
\centerline{\vbox{\offinterlineskip
\hrule
\halign{\vrule height.3cm depth.2cm\quad#\hfil\quad&
\vrule height.3cm depth.2cm\quad\hfil#\hfil\quad&
\vrule height.3cm depth.2cm\quad\hfil#\hfil\quad&
\vrule height.3cm depth.2cm\quad\hfil#\hfil\quad\vrule height.3cm depth.2cm\cr
j$\backslash$k&-1&0&1\cr
\noalign{\hrule}
1&0.69439(-1.29663)&0.66571(-1.19176)&0.68156(-1.27890)\cr
2&0.67586(-1.24041)&0.67407(-1.22997)&0.67034(-1.20080)\cr
3&0.68036(-1.24132)&0.67004(-1.19779)&0.67040(-1.20127)\cr
4&0.66800(-1.17008)&0.66363(-1.07192)&0.66773(-1.16833)\cr
5&0.66669(-1.14729)&0.66720(-1.15800)&0.66378(-1.03678)\cr
6&0.66696(-1.15279)&0.66620(-1.13283)&0.66655(-1.14342)\cr
7&0.66642(-1.13941)&0.66642(-1.13941)&0.66565(-1.10821)\cr
8&0.66642(-1.13941)&0.66600(-1.12451)&0.66601(-1.12484)\cr
9&0.66601(-1.12484)&0.66601(-1.12468)&0.66603(-1.12574)\cr
}\hrule}}
\vskip 1.5truecm
\centerline{\bf Table 4}
\end